\documentclass[prb, twocolumn, showpacs]{revtex4}

\usepackage{bm}
\usepackage{epsfig, amsmath,amssymb}
\input epsf

\begin{document}

\title{Geometry induced charge separation on a helicoidal ribbon}

\author{Victor Atanasov}
 \altaffiliation[Also at]{ Laboratoire de Physique Th\'{e}orique et
Mod\'{e}lisation , Universit\'{e} de Cergy-Pontoise,
 F-95302 Cergy-Pontoise, France }
\affiliation{Institute for Nuclear Research and Nuclear Energy,
Bulgarian Academy of Sciences,  72 Tsarigradsko chaussee, 1784
Sofia, Bulgaria}
 \email{victor.atanasov@u-cergy.fr}

\author{Rossen Dandoloff}
\affiliation{ Laboratoire de Physique Th\'{e}orique et
Mod\'{e}lisation , Universit\'{e} de Cergy-Pontoise,
 F-95302 Cergy-Pontoise, France}
 \email{rossen.dandoloff@u-cergy.fr}

\author{Avadh Saxena}
\affiliation{Theoretical Division and Center for Nonlinear Studies,
Los Alamos National Laboratory, Los Alamos, NM 87545 USA}
\email{avadh@lanl.gov}

\begin{abstract}
We present an exact calculation of the effective geometry-induced quantum
potential for a particle confined on a helicoidal ribbon. This potential leads to
the appearance of localized states at the rim of the helicoid. In this geometry the twist
of the ribbon plays the role of  an effective transverse electric field on the surface 
and thus this is reminiscent of the quantum Hall effect.
\end{abstract}

\pacs{ 02.40.-k, 03.65.Ge, 73.43.Cd}

\maketitle

The interplay of geometry and topology is a recurring theme in physics,
particularly when these effects manifest themselves in unusual
electronic and magnetic properties of materials.  Specifically,
helical ribbons provide a fertile playground for such effects.  Both
the helicoid (a minimal surface) and helical ribbons are ubiquitous
in nature: they occur in biology, e.g. as beta-sheets in protein
strucutres \cite{4}, macromolecules (such as DNA) \cite{5}, and tilted
chiral lipid bilayers \cite{3}.  Many structural motifs of biomolecules
result from helical arrangements \cite{6}: cellulose fibrils in cell
walls of plants, chitin in arthropod cuticles, collagen protein in
skeletal tissue.  Condensed matter examples include screw dislocations
in smectic A liquid crystals \cite{disloc}, certain ferroelectric liquid
crystals \cite{ferro}, and recently synthesized graphene ribbons.  In
particular, graphene M\"obius strips have been investigated for their
unusual electronic and spin properties \cite{Mobius}.  A helicoid to
spiral ribbon transition \cite{1} and geometrically induced bifurcations
from the helicoid to the catenoid \cite{2} have also been studied.

Graphene ribbons can be doped with charges.  In this context, our goal
is to answer the following questions: what kind of an effective quantum
potential does a charge (or electron) experience on a helicoid or a
helical ribbon due to its geometry (i.e., curvature and twist)?  If the
outer edge of the helicoid is charged, how is this potential modified
and if there are any bound states?  Our main findings are: the twist
$\omega$ will push the electrons in vanishing angular momentum state
towards the inner edge of the ribbon and the electrons in non-vanishing
angular momentum states to the outer edge thus creating an inhomogeneous
effective electric field between the inner and outer rims of the helicoidal
ribbon. This is reminiscent of the quantum Hall effect; only here it is
geometrically induced.  We expect our results to lead to new experiments
on graphene ribbons and other related twisted materials where the
predicted effect can be verified.  In a related context we note that
de Gennes had explained the buckling of a flat solid ribbon in terms of
the ferroelectric polarization charges on the edges
\cite{deGennes}.

\begin{figure}[ht]
\begin{center}
   \includegraphics[scale=0.15]{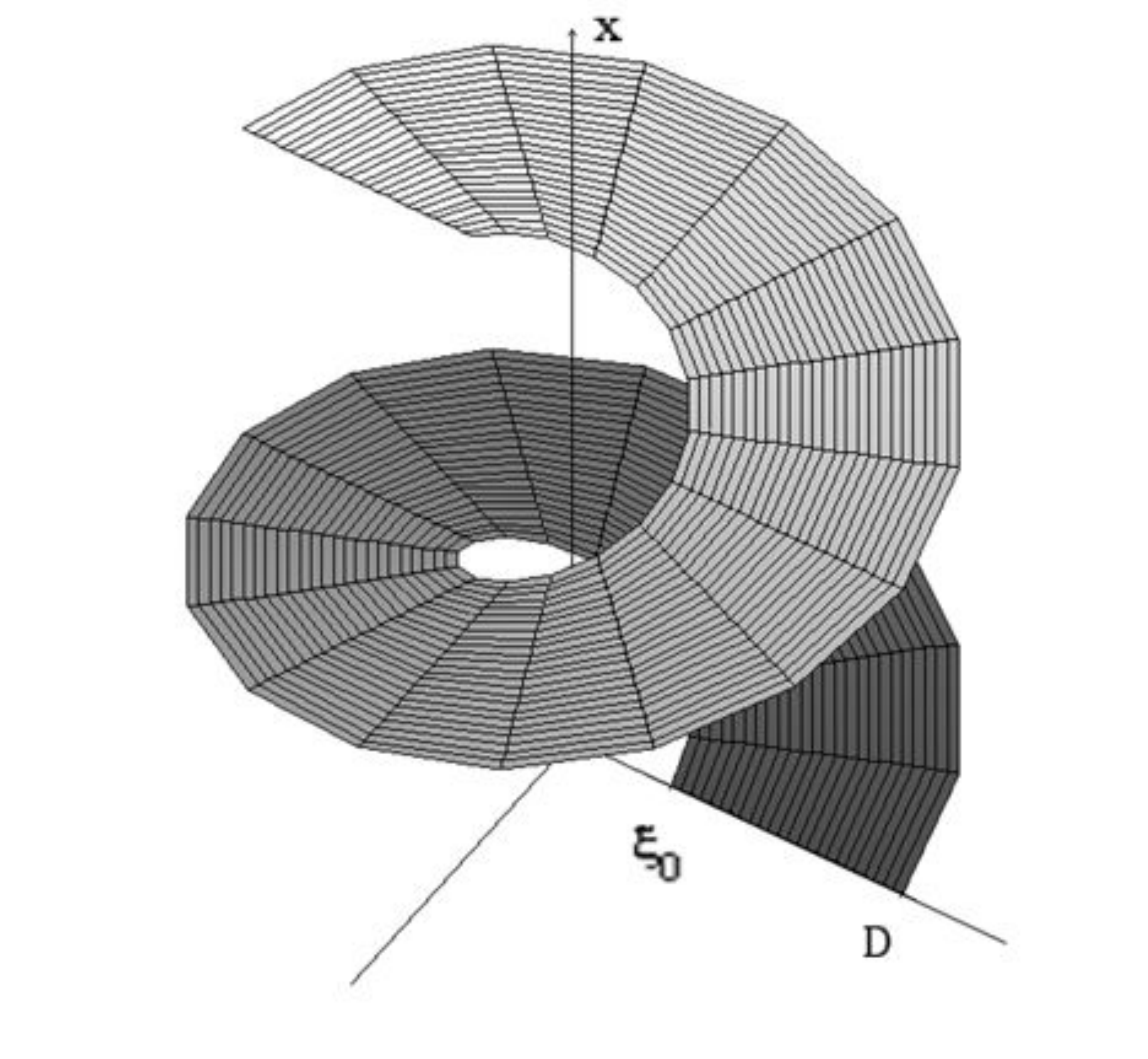}
   \caption{\label{helicoidal} {A helicoidal ribbon with inner radius
$\xi_0$ and outer radius $D$.  For $\xi_0=0$ it becomes a helicoid.  
Vertical axis is along $x$ and the transverse direction $\xi$ is across 
the ribbon.}}
\end{center}
\end{figure}

In order to answer the questions posed above, here we study the helicoidal
surface to gain a broader understanding of the {\it interaction} between
quantum particles and curvature and the resulting possible physical effects.
The properties of {\it free} electrons on this geometry have been
considered before \cite{Dan*04}. The results of this paper are based on
the Schr\"odinger equation for a confined quantum particle on a sub-manifold
of $\mathbb{R}^3$.  Following da Costa \cite{daCosta} an effective potential
appears in the two dimensional Schr\"odinger equation which has the
following form:
\begin{equation}
V_{curv} = - \frac{\hbar^2}{2 m^{\ast}}\left( M^2-K \right) , 
\end{equation}
where $m^{\ast}$ is the effective mass of the particle, $\hbar$ is the
Plank's constant; $M$ and $K$ are the Mean and the Gaussian curvature,
respectively.

To describe the geometry we consider a strip whose inner and outer edges
follow a helix around the $x$-axis (see Fig. \ref{helicoidal} with $\xi_0 
=0$).  The surface represents a {\it helicoid} and is given by the 
following equation:
\begin{equation}
\vec{ r}=x\, \vec{ e}_{x}+ \xi\, [\cos(\omega x)\, \vec{
e}_{y}+ \sin(\omega x)\, \vec{e}_{z}],
\end{equation}
where $\omega = \frac{2\pi n}{L}$, $L$ is the total length of the
strip and $n$ is the number of $2\pi$-twists. Here
$(\vec{e}_x,\vec{e}_y,\vec{e}_z )$ is the usual
orthonormal triad in $\mathbb{R}^3$ and $\xi\in [0, D]$, where $D$ is the
width of the strip. Let ${ d \vec{r} }$ be the line element and the metric is encoded in
\[
|{d \vec{r}}|^2 = (1 + \omega^2\xi^2)dx^2 + d\xi^2 = h_1^2dx^2 +
h_2^2d\xi^2,
\]
where $h_1=h_1(\xi)=\sqrt{1+\omega^2\xi^2}$ and $h_2=1$ are the Lam\'e
coefficients of the induced metric (from $\mathbb{R}^3$) on the strip. 
Here is an appropriate place to add a comment on the {\it helicoidal 
ribbon}, that is a strip defined for $\xi \in [\xi_0, D]$ (see Fig. 
\ref{helicoidal}). All the conclusions still hold true and all of the 
results can be translated using the change of variables
\[
\xi=\xi_0 + s (D-\xi_0), \qquad s\in [0,1].
\]
Here $s$ is a dimensionless variable and one easily sees that for $\xi_0 \to 0$ we again obtain the helicoid.

The Hamiltonian for a quantum particle confined on the ribbon is given by:
\begin{equation}
H=-\frac{\hbar^2}{2m^{\ast}}\frac{1}{h_1}\left[
\left(\frac{\partial}{\partial \xi}h_1\frac{\partial}{\partial
\xi}\right)+ \frac{\partial}{\partial x
}\frac{1}{h_1}\frac{\partial}{\partial x}\right] + V_{curv}.
\end{equation}

Let us elaborate on the curvature-induced potential $V_{curv}.$ Since the
helicoid is a minimal surface $M$ vanishes and we are left with the
following expression
\begin{equation}
V_{curv}= \frac{\hbar^2}{2 m^{\ast}} K  = -\frac{\hbar^2}{2 m^{\ast}} \frac{\omega^2}{[1+\omega^2 \xi^2]^2}.
\end{equation}
Using Gauss' {\it Theorema egregium} \cite{gauss}
the above potential can also be rewritten as
\begin{equation}
V_{curv}= \frac{\hbar^2}{2 m^{\ast}} K  = -\frac{\hbar^2}{2 m^{\ast}}\frac{1}{h_1}\left(
\frac{\partial^2 h_1}{\partial \xi^2}\right)
.
\end{equation}

After rescaling the wave function $\psi \mapsto \frac{1}{\sqrt
h_1}\psi$ (because we require the wave function to be normalized
with respect to the area element $dx d\xi$)
we arrive at the following expression for the Hamiltonian:
\begin{equation}
H=-\frac{\hbar^2}{2 m^{\ast}} \left( \frac{\partial^2}{\partial \xi^2} + \frac{1}{h_1^2}\frac{\partial^2}{\partial x^2}\right) +
V_{eff}(\xi),
\end{equation}
where the effective potential in the (transverse) $\xi$ direction is
given by:
\begin{equation}
V_{eff}(\xi)=-\frac{\hbar^2}{2 m^{\ast}} \left[
\frac{1}{2h_1}\left( \frac{\partial^2 h_1}{\partial \xi^2}\right)
+ \frac{1}{4}\frac{1}{h_1^2}\left(\frac{\partial h_1}{\partial
\xi}\right)^2\right] .
\end{equation}

\begin{figure}[ht]
\begin{center}
   \includegraphics[scale=0.25 ]{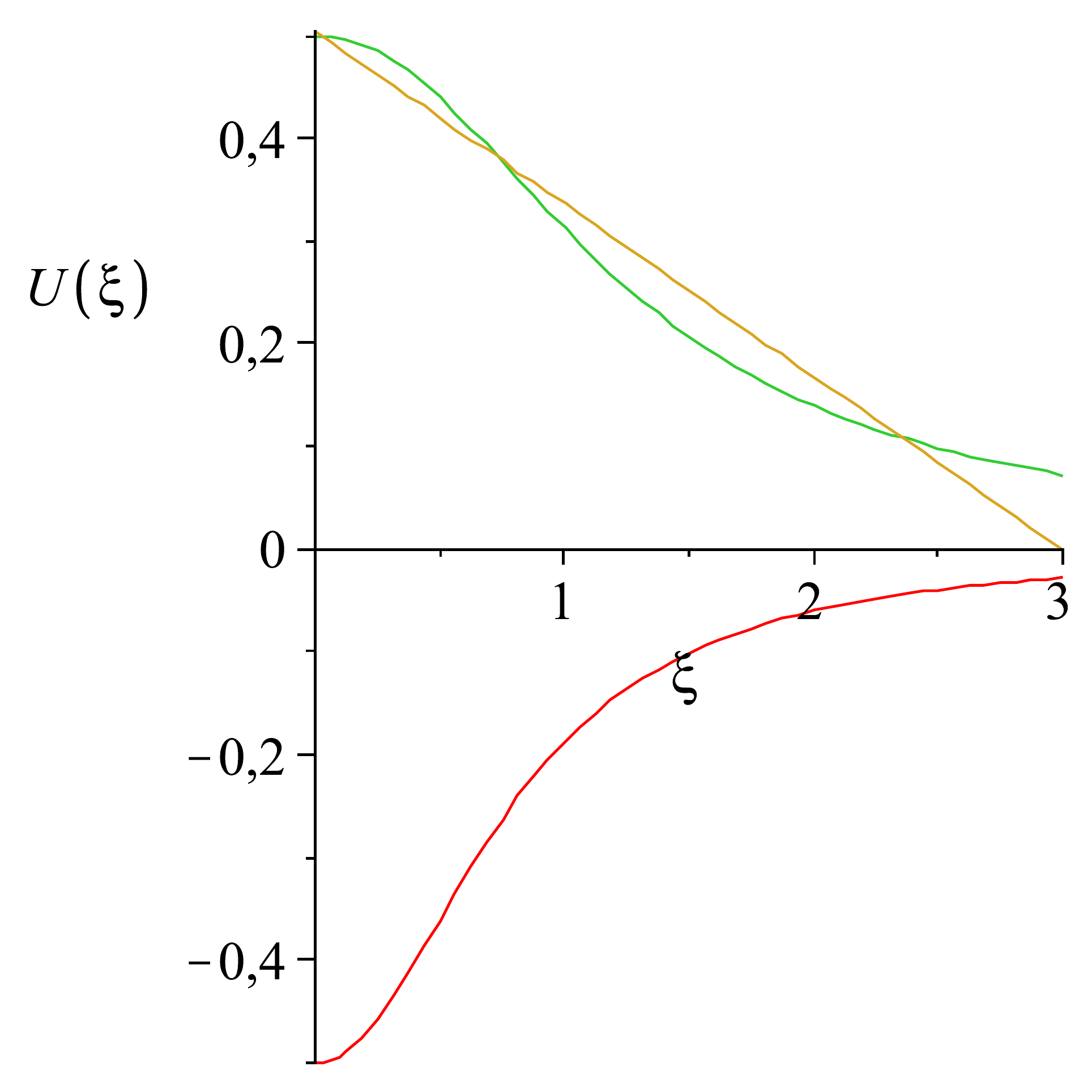}
   \caption{\label{potential} {The behavior of the potential $U(\xi)$ for
$\omega=1$ and $\hbar^2=2m^{\ast}=1.$ Here the red curve corresponds to
$m=0,$ the green curve to $m=1,$ and the yellow line to the approximation
given by Eq. (\ref{approx}).}}
\end{center}
\end{figure}

Note that in bent tubular waveguides \cite{Goldstone} and curved quantum
strip waveguides \cite{Clark} the effective potential is longitudinal. In
the present case there is no longitudinal effective potential. After
insertion of
$h_1=\sqrt{1+\omega^2\xi^2}$ the effective potential becomes:
\begin{equation}
V_{eff}(\xi)=-\frac{\hbar^2}{4 m^{\ast}}\frac{\omega^2}{(1+\omega^2\xi^2)^2}\left[1+\frac{\omega^2\xi^2}{2}\right].
\end{equation}
This effective potential is of pure quantum-mechanical origin
because it is proportional to $\hbar$. Note that this expression
is exact and is valid not just for small $\xi$: here no expansion
in a small parameter has been used.

Next, we write the time-independent Schr\"odinger equation as:
\begin{equation}
\left[-\frac{\hbar^2}{2 m^{\ast}}\frac{\partial^2}{\partial \xi^2} +
V_{eff}(\xi)\right]\psi -
\frac{\hbar^2}{2 m^{\ast}}\frac{1}{h_1^2}\frac{\partial^2 \psi}{\partial
x^2}=E\psi .
\end{equation}
Using the ansatz: $\psi(x,\xi)=\phi (x)f(\xi)$ we split the dependence on the variables
and we get two differential equations:
\begin{equation}\label{SchroX}
-\frac{\hbar^2}{2m^{\ast}}\frac{d^2 \phi(x)}{dx^2}=E_0\phi(x),
\end{equation}
and
\begin{equation}\label{SchroXI}
-\frac{\hbar^2}{2 m^{\ast}}\frac{d^2 f(\xi)}{d\xi^2} + U(\xi) f(\xi)=E f(\xi),
\end{equation}
where
\begin{equation}
U(\xi)=V_{eff}(\xi)+\frac{E_0}{h_1^2(\xi)}.
\end{equation}
With a solution
$\phi(x)=e^{ik_x x}$
of Eq. (\ref{SchroX}) we have
$$E_0=\frac{\hbar^2}{2 m^{\ast}} k_x^2,$$
where $k_x$ is the partial momentum in $x$-direction.
Let us consider here the azimuthal angle around the $x$ axis: $\omega x$ and the angular momentum
along this axis: $
L_x = -  \frac{i \hbar}{\omega} \frac{\partial}{\partial x}.$
 This operator has the same eigenfunctions  $L_x \phi(x)=\hbar m \phi(x)$
as the operator in Eq. (10). The corresponding eigenvalues are $\hbar m.$
We conclude that the momentum $k_x$ is quantized
\[
k_x= m \omega, \qquad m \in \mathbb{N}.
\]
This is not surprising because of the periodicity of the wave function
along $x$. Note that the value of the angular momentum quantum number
determines the direction the electron takes along the $x$ axis either
upward $m>0$ or downward $m < 0.$  This situation is reversed for a
helicoid with opposite chirality.

Equation (\ref{SchroXI}) represents the motion in the direction $\xi$ with
a net potential
\begin{equation}\label{U(xi)}
U(\xi)=- \frac{\hbar^2}{2 m^{\ast}} \frac{\omega^2}{4}\left\{
 \frac{1 -  4 m^2 }{(1+\omega^2\xi^2)} + \frac{1}{(1+\omega^2\xi^2)^2}
\right\} , 
\end{equation}
which is depicted in Fig. \ref{potential}.

This potential is a sum of two contributions, an attractive part:
$\frac{1}{(1+\omega^2\xi^2)^2}$ and a variable part which is repulsive 
for $m \geq 1$ and attractive for $m=0$ (see Fig \ref{potential}). The 
action of this part for $m \neq 0$ qualifies it as a centrifugal potential. 
It pushes a particle to the boundary of the strip. The finite size of the 
width $D$ determines the cut-off of $U(\xi)$ and hence the probability of 
finding the particle is greatest near the rim of the helicoid. Since the 
behavior of the potential  $U(\xi)$ for a particle with $m =0$ qualifies 
it as a quantum anti-centifugal one, it concentrates the electrons around 
the central axis for a helicoid (or the inner rim for a helicoidal ribbon). 
Such anti-centrifugal quantum potentials have been  considered before 
\cite{vic*07}.

The behavior described above can be inferred using the uncertainty 
principle.  Localized states must appear away from the central axis or 
the inner rim. Physically, one may understand the appearance of localized 
states away from the central axis using the following reasoning: for 
greater $\xi$ a particle on the strip will avail  more space along the 
corresponding helix and therefore the corresponding momentum and hence 
the energy will be smaller than for a particle closer to the central axis.

We note that the separability of the quantum dynamics along $x$ and 
$\xi$ directions with different potentials points to the existence of 
an effective mass anisotropy on the helicoidal surface.

For the sake of simplicity let us approximate the potential $U(\xi)$ given
in Eq. (\ref{U(xi)}) (for $m=1$) by a straight line.  The sole purpose of
this approximation is to pinpoint the basic distribution of the probability
density. Assuming it to be linear (see Fig. \ref{potential}) and starting 
from certain $\xi_0=a \ll 1$
\begin{equation}\label{approx}
U_a (\xi) = (D - \xi )\frac{U_0}{D-a}, \qquad U_0=U(\xi=a).
\end{equation}
The value of $a$ can be determined from an area preserving condition
$
\frac12 U_0 (D-a)= \int^D_{\xi_0} U(x)dx, $
where $\xi_0 < D$ is the position from which we evolve the surface. When
dealing with a helicoidal ribbon we must take $\xi_0 \neq 0.$  After
obtaining a result for this case we can easily obtain a result for the
helicoid case by taking the limit $\xi_0 \to 0.$

Next we introduce a characteristic lengthscale $l$ in the problem
\[
l^{-3}=\frac{2 m^{\ast} |U_0|}{\hbar^2 (D-a)}, \qquad \frac{\lambda}{l^2}=\frac{2 m^{\ast}}{\hbar^2} \left( E - \frac{D U_0}{D-a} \right),
\]
where $\lambda$ is a dimensionless energy scale.  After introducing the
dimensionless variable $\zeta=-\lambda - \xi/l$ the Schr\"odinger equation
for the radial part becomes
\begin{equation}\label{Airy}
\frac{d^2 f}{d \zeta^2} - \zeta f(\zeta)=0,
\end{equation}
with the following boundary conditions: $f(-\lambda - \xi_0/l)=f(-\lambda - D/l)=0.$ This form of the equation is valid for $U_0 > 0$ as is the case for $m \neq 0.$

For $m=0$ we have a negative $U_0= - |U_0|$ which requires the introduction of the dimensionless variable $\zeta=-\lambda + \xi/l$ and the corresponding equation is given by (\ref{Airy}), only in this case the  boundary conditions are $f(-\lambda+\xi_0/l)=f(-\lambda + D/l)=0.$

Let us assume that the ratio $D/l \gg 1$ then the solutions, i.e. the wave
functions, of Eq. (\ref{Airy}) coincide with the Airy function, that is
$f(\zeta)= {\rm const \; Ai} (\zeta),$  and the boundary condition
$f(-\lambda \pm \xi_0/l)=0$ (the upper sign corresponds to $m=0$ and the
lower to $m \neq 0$ states) gives the quantized energies
\[
E_n (m) = U_0(m) \frac{D }{D-a} + \left(\lambda_n \pm \frac{\xi_0}{l}\right)\frac{\hbar^2}{2 m^{\ast} l^2},
\]
where $\lambda_n$ are the zeroes of the Airy function ${\rm Ai} (-\lambda_n)=0.$ Let us list the first three of them: $(\lambda_1, \lambda_2, \lambda_3)=(2.338, 4.088, 5.521).$ Here we have taken account of the case when the interior of the helicoid is cut at a distance $\xi_0$  from the axis, that is the ribbon case. The helicoid case is obtained after setting $\xi_0 \to 0.$

For the {\it vanishing angular momentum state} we have $U_0(0) < 0$ and the energy spectrum starts at a negative value (Fig. \ref{potential}), that is we have a {\it bound state.} The probability amplitude has a node at $\xi_0$ in the ribbon case or at the origin for the helicoid case.  The evolution along $\xi$ starts at the corresponding zero of the Airy function and evolves in the positive direction where the Airy function vanishes.  For non-vanishing angular momentum states we have $U_0(m) > 0$ and the energy spectrum is positively valued (Fig. \ref{potential}). The  evolution of the corresponding solutions along $\xi$ starts at the corresponding zero of the Airy function and evolves in the negative direction where the Airy function is oscillatory as one would expect for a confined positive energy spectrum. The observation that the $m  \neq 0$ states at $\xi_0 = \lambda_n l$ have the same energy $E_n (m) = U_0(m) {D }/{(D-a)}$ for all $n$ leads us to believe that this is a particular positive energy oscillatory state whose wavelength fits $D(1-\xi_0/l) \approx D (l/D \gg 1).$

We would like to conclude with the observation that the electric dipole
moment for the $(m=0, \lambda_1)$ bound state (also the ground state for
this geometrical configuration) is non-zero due to the anisotropic
distribution of the probability density along $\xi$. Indeed, suppose we
consider a ribbon doped with a uniform surface charge density $\sigma,$
then the electric dipole vector $\vec{p}=p_x \vec{e}_x + p_{\xi}
\vec{e}_{\xi}$ in the moving coordinate system $(\vec{e}_x,\vec{e}_{\xi},
\vec{e}_3 = \vec{e}_x \times  \vec{e}_{\xi})$  will have non-vanishing $x$
and $\xi$ components:
\begin{eqnarray}
p_x = \frac{Q \pi}{\omega}, &&
p_{\xi} = \frac{2 \pi}{\omega} \sigma l^2 \beta_n,
\end{eqnarray}
where the total charge is $$
Q= \int_0^{2 \pi / \omega} d x' \int_{\xi_0}^D \sigma |\psi(x', \xi')|^2d\xi'$$
and $$
\beta_n= \int_{\xi_0/l}^{D/l \gg 1} \left|{\rm Ai}\left( - \lambda_n \mp \frac{\xi_0}{l} \pm t \right) \right|^2 t d t. $$
Here the upper sign corresponds to $m=0$ and the lower to $m \neq 0$ states. For $\xi_0/l=0.1$ and $D/l=10$ we summarize the values of $\beta_n$ in the following table \\ 

\begin{table}[htdp]
%\caption{default}
\begin{center}
\begin{tabular}{|c|c|c|c|c|}
\hline
& $n=1$ & $n=2$ & $n=3$ & $n=10$ \\
\hline
$m = 0$ & 0.816 & 1.822 & 2.829 & 3.605 \\
\hline
$m \neq 0$ & 2.712 & 2.451 & 2.299 & 1.783\\
\hline
\end{tabular}
\end{center}
%\label{}
\end{table}%

Let us suppose that the outer rim of the helicoid is uniformly charged or
there is a uniformly charged wire going through the core, then this will
create an accelerating electric field term in the effective potential
$U(\xi),$ that is $ U^e(\xi)=U(\xi)+e \mathcal{E}\xi.$ The dynamics is
still separable. In the cup-shaped potential $U^e$ the electrons will be
found with the greatest probability where the potential has a minimum.
This means that the extra charge on the helicoid will concentrate in a
strip around the value of $\xi_{min}$, i.e. a solution to $d U^e / d \xi
= - e \mathcal{E}.$

Application of an electric or magnetic field along the $x$-axis would
nontrivially affect the motion of electrons on the surface--this problem
will need to be studied numerically.  It would be very interesting to
observe the predicted effect in graphene ribbons or helicoidal ribbons
synthesized from a semiconducting material.

Our main findings can be summarized as follows: the twist $\omega$ will
push the electrons with $m \neq 0$  ($m=0$) towards the outer (inner)
edge of the ribbon and {\it create an effective electric field between
the central axis and the helix, the latter representing the rim of the
helicoid.}  Instead of a helicoidal ribbon, if we consider a cylindrical
helical ribbon then both the curvature and torsion are constant and the
effective potential is quite simple.  We expect our results to motivate
new low temperature ($T < \hbar^2/k_B 2 m^{\ast} l^2,$ where $k_B$ is
Boltzmann's constant) experiments on twisted materials.

This work was supported in part by the U.S. Department of Energy.

\end{document}